\documentclass[prd,nofootinbib,superscriptaddress,twocolumn]{revtex4}
\usepackage{amstext,amsmath,amssymb,amsfonts,bbm}
\usepackage{euscript}
\usepackage[usenames, dvipsnames]{color}
 \usepackage{epsfig}

\usepackage{ulem}

\def\be{\begin{equation}}
\def\ee{\end{equation}}
\def\bes{\begin{eqnarray}}
\def\ees{\end{eqnarray}}
\def\bay{\begin{array}}
\def\ear{\end{array}}

\def\pad{{\partial}}
\def\1{{{\mathbbm 1}}}

\def\sg{\textsl{g}}

\begin{document}
\title{ Do event horizons exist? }
\author{Valentina Baccetti}
\email{valentina.baccetti@mq.edu.au}
\affiliation{Department of Physics and Astronomy, Macquarie University, Sydney NSW 2109, Australia}
\author{Robert B. Mann}
\email{rbmann@uwaterloo.ca}
\affiliation{Department of Physics and Astronomy, University of Waterloo, Waterloo, Ontario N2L 3G1, Canada }
\affiliation{Perimeter Institute for Theoretical Physics, Waterloo, Ontario N2L 2Y6, Canada}
\author{Daniel R. Terno}
\email{daniel.terno@mq.edu.au}
\affiliation{Department of Physics and Astronomy, Macquarie University, Sydney NSW 2109, Australia}

\begin{abstract}

  Event horizons are the defining feature of  classical black holes.  They are  the key ingredient of the information loss paradox which, as paradoxes in quantum foundations, is built  on a combination of  predictions of quantum theory and counterfactual classical features: neither  horizon formation nor its crossing by a test body is observable. Furthermore, horizons are
unnecessary for the production of  Hawking-like radiation.  We demonstrate that when this radiation is taken into account it  prevents horizon crossing/formation in a large class of models.  We  conjecture that horizon avoidance is  a general feature of collapse.
The non-existence of event horizons  dispels the paradox, but opens up important questions  about thermodynamic properties of the resulting objects and   correlations between different degrees of freedom.

\bigskip

\bigskip

\bigskip

\end{abstract}

\date{\today}

\maketitle

 Event horizons   ---  null surfaces   bounding   spacetime  regions from which  signals cannot escape  --- are the defining  feature of  classical black holes \cite{fn:98,ast-bh}. They are built into quantum   black hole models,  beginning with  semi-classical quantum field theory on curved backgrounds  and reaching to loops and strings \cite{fn:98,kiefer:07}.  The foremost prediction of  the semi-classical theory is emission of  Hawking radiation \cite{h:74}. Its original derivation  relied on the existence of an horizon.  Identification of surface gravity at the horizon with temperature  \cite{fn:98,wald:01} completed  black hole thermodynamics,  but  inadvertently formented the  information loss paradox, perhaps  the longest-running controversy in theoretical physics  \cite{wald:01,modern,zeh}.

  Event horizons provide the most obvious causal decomposition  of  spacetimes,  and motivate the tensor product structure of the Hilbert spaces of gravity and matter states. When   matter collapses into a black hole, it   evaporates  via Hawking radiation within a finite time. If quantum correlations between the inside and outside of the black hole horizon are not restored during the evaporation, then this evolution of low-entropy matter into  high-entropy radiation implies information loss \cite{wald:01}. This problem is described as a paradox  \cite{fn:98,modern,bht:17} because a combination of information-preserving   theories --- quantum mechanics and general relativity --- ostensibly leads to a loss of information.


According to a distant observer (Bob)  both formation of black holes from collapsing matter and their absorption of test particles
  take an infinite amount of time. Bob's clock indicates that  approach to the Schwarzschild radius $r_\sg=2M$ is exponentially slow.  
 Still, after at most a few dozen multiples of light-crossing time  $r_\sg$, he cannot receive signals from Alice (an observer  co-moving with the matter {who is initially at (or near) its edge}); the red-shift requires that the energy of any such detected signal is greater than the mass of  the black hole.  Hence for external observers  formation of an event horizon or its crossing are in principle  unobservable \cite{visser:14}.

Consequently co-moving Alice's  experiences, like crossing the Schwarzschild radius, become counterfactual. Her clock readings
indicate that the collapsing matter or  test particle cross $r_\sg$  in a finite time $\Delta\tau$,  proving that   coordinate systems employing   time at infinity are geodetically incomplete. As Alice cannot communicate her clock readings to Bob, this is an experimentally unverifiable consequence of the formalism of general relativity. Nevertheless, finite proper crossing time
promotes the event horizon  and, by extension, the quantum states associated with black hole horizon and interior, from  convenient mathematical concepts to physical  entities.

  Counterfactual reasoning, not illegitimate in itself,   is nevertheless responsible for  many paradoxes of quantum mechanics, particularly when  intuitively reasonable features of classical theory are combined with elements of quantum formalism \cite{peres:95,imt:15}. The information loss paradox is built from the same components.

Event horizon was employed in the original derivation of Hawking radiation in a static spacetime \cite{fn:98,modern}, but can be dispensed with in a dynamical spacetime of  collapsing matter.  Numerical studies  and  analytic  results, such as \cite{haj:86,acvv:01,blsv:06,vsk:07,kmy:13},  establish  the  existence of a pre-Hawking radiation.  Once its effects are taken into account, the Schwarzschild radius becomes unattainable.

 To demonstrate this, consider  a massive thin shell, whose classical physics  is well-understood \cite{poisson}. Spacetime inside the shell is flat, and we may use here standard Minkowski coordinates $(t_-,r_-)$. The exterior geometry is described by the Schwarzschild metric
 \be
 ds^2= -f(r) dt^2+f(r)^{-1}dr^2 + r^{2} {d\Omega}, \label{met2}
 \ee
where  $f(r)=1-r_\sg/r$.   The shell's trajectory is parameterized by Alice's co-moving proper time  $\big(T(\tau), R(\tau)\big)$ and $\big(T_-(\tau), R_-(\tau)\big)$, in  exterior and interior coordinates respectively.  The dynamics is obtained with the help of the two junction conditions \cite{poisson}.

 The first   junction condition requires the induced shell metric to be the same on both  {of} its sides, implying
 the identification $R_-(\tau)\equiv R(\tau)$. Discontinuity of the extrinsic curvature is described by the second junction condition, and once supplemented by the surface stress-energy tensor it leads to the equations of motion of the shell. For pressureless dust the equations are simple enough to have an analytic solution $\tau=\tau(R)$ that demonstrates a finite horizon crossing time $\tau(r_\sg$).

There is no ready-made prescription that provides  the expectation value of the stress-energy tensor that  feeds into the Einstein equations and self-consistently produces a metric. However, such quantum effects should be expressible in terms of an expectation value of the stress energy tensor and  {a corresponding consistent
 metric of the form  \cite{visser:14}}
\be\label{sch-co}
ds^2=-k(t,r)^2f(t,r)dt^2+f(t,r)^{-1}dr^2+r^2d\Omega,
\ee
 {that describes spacetime outside the evaporating shell, }
where the functions $k(t,r)$  and $f(t,r)=1-C(t,r)/r$ satisfy certain mild restrictions.

Even without knowing the explicit form of these functions it is possible to draw conclusions about the shell's dynamics. 
We assume that (i) $0\leq C<\infty$ with $C(t,r)>0$  for $t<t_E<\infty$, where $t_E$ is the evaporation time,  and   $\pad C/\pad  t<0$  as long as  $C>0$, ensuring finite positive gravitational mass, positive energy density and  positive flux at infinity; (ii) $k(t,r)$ is continuous; (iii) the metric has only one coordinate singularity, namely a   surface given by $f(t,r)=0$. This surface is located at the Schwarzschild radius   $r_\sg(t)$, that  is   implicitly defined by $r_\sg\equiv C(t,r_\sg)$.

 By monitoring the gap between the shell and the Schwarzschild radius \cite{bmt:16},
\be\label{gap}
x(\tau):=R(\tau)-r_\sg\big(T(\tau)\big),
\ee
we discover  how evaporation modifies the classical shell dynamics.

The   above assumptions imply  \cite{bmt:17}
\be
C(t,r)=r_\sg(t)+w(t,r)\big(r-r_\sg(t)\big), \label{sch-co-C}
\ee
where the functions $w(t,r)$ and $r_\sg(t)$ are constrained by   (i)-(iii). In particular, they satisfy the inequalities $w(t,r_\sg)<1$  and $dr_\sg/dt<0$.
Since massive particles move on the time-like trajectories,   components of the shell's four-velocity satisfy
\be
\dot{T}=\frac{\sqrt{F+\dot R^2}}{|K|F}>\frac{|\dot{R}|}{|K|F},
\ee
where $K=k\big(T(\tau),R(\tau)\big)$.
Close to the Schwarzschild radius we expand $F$ and $\dot T$  {in  powers of $1/x$   obtaining  \cite{bmt:17}}
\be
\dot{x}>\dot R(1-\epsilon_*/x), \label{xdotgen}
\ee
where the scale of a potential horizon avoidance \cite{bmt:17} is set by
\be
\epsilon_*=\frac{1}{|K|}\frac{r_\sg}{1-W}\left|\frac{d r_\sg}{d T}\right|,
 \label{epstasch}
\ee
 with $W(T):=w\big(T,r_\sg(T)\big)$.

  The gap \eqref{gap} decreases only as long as $\epsilon_*<x$. {As a result, while it decreases, $R>r_\sg+\epsilon_*$.} {If} the distance between the shell and the Schwarzschild radius is reduced to  {$x=\epsilon_*$}, it cannot decrease any further.
{In either case from distant Bob's viewpoint  the shell is still stuck within a slowly changing coordinate distance $\epsilon_*$ from the slowly receding Schwarzschild radius. For   co-moving Alice the collapse accelerates, but never enough to lead to}  horizon crossing.

Using the outgoing Vaidya metric as an example  \cite{bmt:17} makes is straightforward to show that Alice will see  the evaporation rate $\dot C$ vanish if and only if Bob does. In such a case  $\epsilon_*\rightarrow 0$, and an horizon forms. During the ongoing evaporation, since the gap $x$ never vanishes and $\dot R$ increases but remains finite for finite $x$, Alice will see the radiation flux  grow to some maximal value
contingent on the properties of the shell.  This could be regarded as a firewall with a natural upper cut-off \cite{bmt:17}, providing a possible realization of the firewall acting via `internal conversion' \cite{zeh}.

{In other words, $r_\sg$ is a hypothetical surface that  the shell gets very close to  but never crosses. Neither trapped surface, nor horizon nor singularity ever form.} The distance $\epsilon_*\propto C^{-1}$ \cite{bmt:16} grows as the shell evaporates.  If the   radiation stops at some point, then
the remaining shell will collapse into a black hole.

 We can use this result to understand how  emission of the pre-Hawking radiation modifies     Oppenheimer-Snyder {collapse \cite{bmt:17}.  The simplest scenario involves an initially uniform distribution of pressureless dust \cite{os:39,poisson,ast-bh} that is  also not interacting with the pre-Hawking radiation field.  Without evaporation the interior metric  is that  of the Friedmann-Lema\^{i}tre-Robertson-Walker closed universe,  and its parameters are obtained by matching with the Schwarzschild metric outside.

Assuming  the   dynamics  modified by the pre-Hawking radiation  is such that   individual layers of dust do not cross, the initial parametrization by the radial coordinate $\chi$ fixes their co-moving radius.  Each layer can be parameterized as $\big(T_\chi(\tau), R_\chi(\tau)\big)$, where $\tau$ is common co-moving time for the matter.  Each particle still moves along  a geodesic, and at each layer the metric is of the form of Eq.~\eqref{sch-co} with the appropriate functions, such as $f_\chi\big(T_\chi(\tau),R_\chi(\tau)\big)$. The Schwarzschild radius that any given layer could potentially cross is $r_\sg^\chi\big(T_\chi(\tau)\big)$; the previous discussion ensures that this cannot happen in finite $\tau$ \cite{bmt:17}.

Our results are consistent with the existence of super-compact objects (``black holes without horizon"), such as proposed in \cite{mersini:14,
Saravani:2012is} or fuzzballs \cite{mathur:05} on the one hand, and with general arguments that the putative event horizon is destroyed by quantum effects \cite{brust:14,visser:14}  on the other. It is reasonable to conjecture that horizon formation does not take place for collapsing objects in
any configuration, provided that the pre-Hawking radiation is taken into account and  appropriate energy conditions are assumed.

The horizon avoidance scale $\epsilon_*$ is sub-Planckian, therefore one would expect that from the point of view of a distant observer (Bob), the predictions of this model are indistinguishable from pure classical collapse. However, as indicated by the results in \cite{cardoso:16}, the qualitatively different nature of the resulting objects could potentially be detected via gravitational waves.

If event horizons are absent, what of  the information loss paradox?
The standard Penrose diagram  illustrating   black hole creation and evaporation  includes an event horizon, its crossing by some matter, and eventual evaporation --- this is
clearly inapplicable \cite{bht:17}. Approaches endeavouring to preserve unitary dynamics
that are based on the analysis of matter alone require
both an horizon and a singularity and so are  not applicable as well.
Indeed, our results
 strengthen the point of view  that fully quantized joint gravity-matter dynamics must
have unitary time evolution, particularly for systems that have
a well-defined classical Hamiltonian, and so there cannot be
any overall information loss \cite{bht:17,k:98,ht:10}.  The information loss paradox then goes the way other quantum paradoxes.

Even if there is no paradox,  important unanswered questions remain. How does entanglement (and more general types
of quantum correlations) get distributed between the tripartite
system of gravity/early modes/late modes?  Bekenstein-Hawking black hole entropy $S_{BH}=A/4=\pi r_\sg^2$ is a quarter of the horizon area in Planck units. If event horizons do not correspond to   asymptotically reachable states of  collapsing matter, what are the thermodynamic properties  of the resulting ultra-compact objects?


\medskip
\textbf{Acknowledgments} The authors would like to thank Bernard S. Kay, H. Dieter Zeh, Victor Cardoso, and Paolo Pani for useful discussions. VB is supported by the Macquarie Research Fellowship scheme.

\end{document}